\documentstyle[11pt]{article}
\begin{document}
\begin{center}
{\Large {\bf Electromagnetic masses of the mesons and the generalization of
Dashen's theorem}}\\[5mm]
Dao-Neng Gao\\
{Center for Fundamental Physics, University of Science and
Technology of China\\
Hefei, Anhui 230026, People's Republic of China}\\[2mm]
Mu-Lin Yan\\
{Chinese Center for Advanced Science and Technology(World Lab)\\
P.O.Box 8730, Beijing, 100080, People's Republic of China}\\
{Center for Fundamental Physics,
University of Science and Technology of China\\
Hefei, Anhui 230026, People's Republic of China}\footnote{mailing address}
\end{center}
\vspace{8mm}
\begin{abstract}
\noindent
In the framework of $U(3)_L\times U(3)_R$ chiral field theory, Dashen's theorem
is reexamined, and the well-known result of $m_{\pi^\pm}^2-m_{\pi^0}^2$ obtained
by Das, Guralnik, Mathur, Low, and Young is reproduced. We find that
Dashen's theorem, which automatically holds for pseudoscalar mesons in this
theory, can be generalized to the sector of axial-vector mesons, however,
fails for the sector of vector mesons.
\end{abstract}

\vspace{4mm}
{\bf PACS} number(s): 13.40.Dk, 11.30.Rd

\newpage
\section{Introduction}
\par
In the recent years, studies of electromagnetic masses of mesons,
Dashen's theorem \cite{Da} and its violation have attracted much
attention \cite{DHW,JB,KK,RU,DP,GYL,Mou}.
Dashen's theorem, which states that the square electromagnetic mass differences
between the charge pseudoscalar
mesons and their corresponding neutral partners are equal in the chiral
SU(3) limit, is expressed as follows,
\begin{eqnarray}
& &(m_{\pi^\pm}^2-m_{\pi^0}^2)_{EM}=(m_{K^\pm}^2-m_{K^0}^2)_{EM},\nonumber \\
& &(m_{\pi^0}^2)_{EM}=0,\;\;\;\;(m_{K^0}^2)_{EM}=0.
\end{eqnarray}
The subscript EM denotes electromagnetic mass.
There are three meson-octets $0^-$(pseudoscalar mesons), $1^-$(vector mesons), and $1^+$
(axial-vector mesons) which belong to the same representation
of $SU(3)\subset U(3)$,
but have different spin or
parity. Thus, a natural question aroused here is
whether
Dashen's theorem(which holds for pseudoscalar-octet mesons) could be
generalized to
the vector-octet and axial-vector-octet mesons or not. In this
present paper, by employing
$U(3)_L\times U(3)_R$ chiral theory \cite{Li1,Li2}, it will be shown that the
generalization to axial-vector mesons is valid, or, in the chiral SU(3)
limit, 
\begin{eqnarray}
& &(m_{a^\pm}^2-m_{a^0}^2)_{EM}=(m_{K_1^\pm}^2-m_{K_1^0}^2)_{EM},\nonumber \\
& &(m_{a^0}^2)_{EM}=0,\;\;\;\;(m_{K_1^0}^2)_{EM}=0.
\end{eqnarray}
However, similar generalization fails for vector mesons.
The latter is of the same conclusion as one given by
Bijnens and Gosdzinsky \cite{JG}.

$U(3)_L\times U(3)_R$ chiral theory of pseudoscalar, vector, and
axial-vector mesons provides a unified description of meson physics
in low energies. This theory has been extensively investigated
in Refs.\cite{Li1,Li2,Li3,GYL,WY,Li4}, and its predictions are in good
agreement with the experimental data.
Vector meson dominance(VMD) \cite{JS} in the meson physics is a
natural consequence of this theory instead of an input. This means that
the dynamics of
the electromagnetic interactions of mesons has been introduced and
established naturally. Therefore, the present theory makes it possible to
evaluate the electromagnetic self-energies of these low-lying mesons
systematically. According to this pattern, for example, we can work out 
the well-known result of $(m_{\pi^\pm}^2-m_{\pi^0}^2)_{EM}$ given
by Das et al \cite{DGMLY}(to see Sec. 2), which serves as leading order
in our evaluations.
This indicates that our pattern
evaluating the electromagnetic self-energies of mesons is consistent with the
known theories in the pion case under the lowest energies limit.
Since the dynamics of mesons, including pseudoscalar, vector and 
axial-vector, is described in the present theory, the method calculating 
EM-masses in this paper is legitimate not only for $\pi$ and $K$ mesons,
but also for $a_1$, $K_1$, $\rho$ and $K^*$ mesons. Thus, both
checking the Dashen's argument of eq.(1) in the
framework of the effective quantum fields theory
and investigating its generalizations mentioned above become
practical.

The basic lagrangian of this chiral field theory is (hereafter we use the
notations in Refs.\cite{Li1,Li2})
\begin{eqnarray}
\lefteqn{{\cal L}=\bar{\psi}(x)(i\gamma\cdot\partial+\gamma\cdot v
+e_0Q\gamma\cdot A+\gamma\cdot a\gamma_{5}
-mu(x))\psi(x)}\nonumber \\
& &+{1\over 2}m^{2}_{1}(\rho^{\mu}_{i}\rho_{\mu i}+
\omega^{\mu}\omega_{\mu}+a^{\mu}_{i}a_{\mu i}+f^{\mu}f_{\mu})\nonumber\\
& &+{1\over2}m^2_2(K_{\mu}^{*a}K^{*a\mu}+K_1^{\mu}K_{1\mu})\nonumber \\
& &+\frac{1}{2}m^2_3(\phi_{\mu} \phi^{\mu}+f_s^{\mu}f_{s\mu})
+{\cal L}_{EM}
\end{eqnarray}
where $u(x)=exp[i\gamma_{5} (\tau_{i}\pi_{i}+\lambda_a K^a+\eta
+\eta^{\prime})]$($i$=1,2,3 and $a$=4,5,6,7), $a_{\mu}=\tau_{i}a^{i}_{\mu}+
\lambda_a K^a_{1\mu}+(\frac{2}{3} +\frac{1}{\sqrt{3}} \lambda_8)f_{\mu}+(
\frac{1}{3}-\frac{1}{\sqrt{3}} \lambda_8)f_{s\mu},
v_{\mu}=\tau_{i}\rho^{i}_{\mu}+\lambda_a K_{\mu}^{*a}+(\frac{2}{3}+ \frac{1}{
\sqrt{3}}\lambda_8)\omega_{\mu}+(\frac{1}{3}- \frac{1}{\sqrt{3}}%
\lambda_8)\phi_{\mu}$, $A_\mu$ is the photon field, $Q$ is the electric
charge operator of $u$, $d$ and $s$ quarks, $\psi$ is
quark-fields, $m$
is a parameter related to the quark condensate, and ${\cal L}_{EM}$ is the
kinetic lagrangian of photon fields.
 Here, the meson-fields are bound states of quarks,
and they are not fundamental fields. The effective lagrangian for mesons
is derived by performing path integration over quark fields. This
treatment naturally leads to VMD.
Following Refs.\cite{Li1,Li2},
the interaction lagrangians of neutral vector meson fields and photon
fields read
\begin{eqnarray}
{\cal L}_{\rho \gamma}=-{e \over f_\rho} \partial_\mu
    \rho_\nu^0 (\partial^\mu A^\nu -\partial^\nu A^\mu),\nonumber \\
{\cal L}_{\omega \gamma}=-{e \over f_\omega} \partial_\mu
    \omega_\nu (\partial^\mu A^\nu -\partial^\nu A^\mu),\nonumber \\
{\cal L}_{\phi \gamma}=-{e\over f_\phi}  \partial_\mu
    \phi_\nu (\partial^\mu A^\nu-\partial^\nu A^\mu).
\end{eqnarray}
The direct couplings of photon to other mesons are constructed through the
substitutions 
\begin{eqnarray}
\rho_\mu^0\rightarrow \frac{e}{f_\rho} A_\mu,\;\;\;
\omega_\mu\rightarrow \frac{e}{f_\omega} A_\mu,\;\;\;
\phi_\mu\rightarrow \frac{e}{f_\phi} A_\mu.
\end{eqnarray}
where 
\begin{eqnarray}
\frac{1}{f_\rho}=\frac{1}{2}g,\;\;\;\frac{1}{f_\omega}=\frac{1}{6}g,
\;\;\;\frac{1}{f_\phi}=-\frac{1}{3\sqrt{2}}g.
\end{eqnarray}
$g$ is a universal coupling constant in this theory. It can be determined by
taking the experimental values of $f_\pi$, $m_\rho$ and $m_a$ as
inputs \cite{Li1,Li2,GYL}. Thus, all interaction lagrangians between
mesons and
photon($\gamma$), ${\cal L}_i (\Phi, \gamma,...)|_{\Phi=\pi, a, v...}$ are
obtained.

Using ${\cal L}_i (\Phi,\gamma,...)|_{\Phi=\pi, a, v,...}$, 
we can evaluate the following S-matrix 
\begin{equation}
S_{\Phi}=\langle \Phi |T{\rm exp}[i\int d^4x {\cal L}_i
(\Phi,\gamma,...)]-1|\Phi\rangle |_{\Phi=\pi, a, v,...}. 
\end{equation}
On the other hand $S_\Phi$ can also be expressed in terms of the effective
lagrangian of $\Phi$ as 
$$
S_\Phi=\langle \Phi |i\int d^4 x {\cal L}_{{\rm eff}} (\Phi) | \Phi \rangle. 
$$
Noting ${\cal L}={\frac{1}{2}}\partial_\mu\Phi\partial^\mu \Phi-{\frac{1}{2}}
m_{\Phi}^2 \Phi^2$, then the electromagnetic interaction correction to the
mass of $\Phi$ reads 
\begin{equation}
\delta m_\Phi^2 ={\frac{2iS_\Phi }{\langle \Phi |\Phi^2 | \Phi \rangle }}. 
\end{equation}
where $\langle \Phi |\Phi^2 |\Phi \rangle = \langle \Phi |\int d^4 x
\Phi^2(x) |\Phi \rangle $. Thus, all of virtual photon contributions to mass
of the mesons can be calculated systematically.

The paper is organized as follows. In Sec. 2, we shall reexamine Dashen's
theorem for pseudoscalar mesons in the framework of the present theory.
In Secs. 3 and 4, the generalization of this theorem to axial-vector and
vector mesons is studied respectively. Finally, we give the summary and
conclusions.

\section{To reexamine Dashen's theorem in $U(3)_L\times U(3)_R$
chiral theory of mesons}
\par

Due to VMD(eq.(4)), the interaction lagrangians
which contribute to electromagnetic mass of the mesons has to contain the
neutral vector meson fields $\rho^0$, $\omega$ and $\phi$. When
the calculations are of $O(\alpha_{em})$ and one-loop, there are two sorts
of vertices contributing to electromagnetic self-energies of pseudoscalar
mesons: four points vertices and three points vertices. The former must
be the coupling of two pseudoscalar fields and two neutral vector mesons
fields, and the latter must be the interaction of a pseudoscalar field to a
neutral vector meson plus another field.

Thus, from Refs.\cite{Li1,Li2}, the interaction lagrangians contributing
to
electromagnetic mass of pseudoscalar mesons($\pi$ and $K$) via VMD in the
chiral limit read
\begin{eqnarray}
& &{\cal L}_{\rho \rho \pi \pi}={4F^2\over g^2 f_\pi^2}(\rho_\mu^0
\rho^{0\mu}+{1 \over 2\pi^2 F^2 }\partial_\nu\rho_\mu^0 \partial^\nu\rho^{0\mu})
  \pi^+\pi^-,\\
& &{\cal L}_{\rho \pi a}={2i\gamma F^2 \over f_\pi g^2}\rho_\mu^0
\pi^+ (a^{-\mu}-{1 \over 2\pi^2 F^2}\partial^2 a^{-\mu})+h.c.,  \\
& &{\cal L}_{K^+K^-v v}=\frac{F^2}{f_\pi^2 g^2}\{
 (\rho^{0}_{\mu}+v_\mu^8)^2+\frac{1}{2\pi^2
F^2}(\partial_\nu\rho^{0}_{\mu}
 +\partial_\nu v_\mu^8)^2\}K^+K^{-},\\
& &{\cal L}_{K^{\pm} K_1^{\pm} v}=\frac{i\gamma F^2}{g^2 f_\pi}(\rho_\mu^0+v_\mu^8)
K^+ (K_1^{-\mu}-\frac{1}{2\pi^2 F^2}\partial^2 K_1^{-\mu})+h.c.,\\
& &{\cal L}_{K^0\bar{K}^0 v v}=\frac{F^2}{f_\pi^2 g^2}\{
 (-\rho^0_\mu+v_\mu^8)^2+\frac{1}{2\pi^2 F^2}(-\partial_\nu\rho^0_\mu
 +\partial_\nu v_\mu^8)^2\}K^0\bar{K}^0,\\
& &{\cal L}_{K^{0} K_1^{0} v}=\frac{i\gamma F^2}{g^2 f_\pi}(-\rho_\mu^0+v_\mu^8)
K^0 (\bar{K}_1^{0\mu}-\frac{1}{2\pi^2 F^2}\partial^2 \bar{K}_1^{0\mu})+h.c.
\end{eqnarray}
where 
\begin{eqnarray*}
\pi^\pm=\frac{1}{\sqrt{2}}(\pi^1\pm i\pi^2),\;\;\; a^\pm=\frac{1}{\sqrt{2}}
(a^1\pm i a^2),\\
K^\pm=\frac{1}{\sqrt{2}}(K^4\pm i K^5),\;\;\; K^0(\bar{K}^0)=
\frac{1}{\sqrt{2}}(K^6\pm i K^7),\\
K_{1\mu}^\pm=\frac{1}{\sqrt{2}}(K^4_{1\mu} \pm i K^5_{1\mu}),\;\;\;
K_{1\mu}^0(\bar{K}_{1\mu}^0)=\frac{1}{\sqrt{2}}(K^6_{1\mu}\pm i
K^7_{1\mu}).
\end{eqnarray*}
with 
\begin{eqnarray}
& &F^2={f_\pi^2 \over 1-\frac{2c}{g}},\;\;\;c={f_\pi^2\over 2 g m_\rho^2},\\
\nonumber
& &\gamma=(1-\frac{1}{2\pi^2 g^2})^{-1/2}.
\end{eqnarray}
where $v$ denotes the neutral vector mesons $\rho^0$, $\omega$ and $\phi$,
$v_\mu^8=\omega_\mu-\sqrt{2}\phi_\mu$. $
f_\pi$ is pion's decay constant, and $f_\pi=0.186GeV$.
Here, we neglect the contributions to
electromagnetic mass of pions or kaons which are proportional to $m_\pi^2$
or $m_K^2$, because we are interested in the case of chiral limit.

Note that there are no contributions to electromagnetic masses of $\pi^0$ in the
chiral limit. This means 
\begin{equation}
({m_{\pi^0}^2})_{EM}=0,\;\;\;\;{for\;\; massless\;\; quark.} 
\end{equation}

It can be see from eqs.(9-14) that the kaon's lagrangian is different from
pion's. All three neutral vector resonances $\rho^0$,$\omega$ and $\phi$
join kaon's dynamics there, but only $\rho^0$ joins pion's. In general,
because of this difference the kaon's electromagnetic mass due to VMD
must be different from the pion's. Taking a glance at this
circumstance,
it seems to be impossible to expect the $K^0$'s electromagnetic mass
vanishes like eq.(16) for $\pi^0$'s. Fortunately, however, after a
careful consideration it will be shown below that $K^0$'s electromagnetic
mass does vanish in the chiral SU(3) limit in our formalism.
The point is that in the chiral SU(3)
limit, $m_u=m_d=m_s=0$ and $m_\rho=m_\omega=m_\phi$, the 
SU(3)-vector-meson symmetry and VMD make the total contribution to
$(m_{K^0}^2)_{EM}$ vanished.
 
Let us show this point precisely now. The lagrangians contributing to
electromagnetic masses of $K^0$ (eqs.(13)(14)) are
different from $K^\pm$'s (eqs.(11)(12)).
This difference is due to
the structure constants of SU(3) group: $f_{345}=-f_{367}=\frac{1}{2},
f_{458}=f_{678}=\frac{\sqrt{3}}{2}$. 
In the $K^0$'s lagrangians, eqs.(13)(14), the neutral vector mesons
emerge in combination forms: $(-\rho^0_{\mu}+v^8_{\mu})$ or 
$(-\partial_\nu \rho^0_\mu+\partial_\nu v_\mu^8)$.
Note that in the calculations of electromagnetic masses of pseudoscalar
mesons, the vector meson fields( $\rho
$, $\omega$ and $\phi$) and axial-vector meson fields ($a_1$ and $K_1(1400)$)
in the above lagrangians must be abstracted into propagators in the
corresponding
S-matrices. Therefore, from eqs.(11)-(14) and SU(3) symmetry limit, $
m_\rho=m_\omega=m_\phi$, it can be sure that $v_\mu^8$ is actually
equivalent
to $\rho_\mu^0$ in the calculations of electromagnetic masses of the
mesons.
Thus, the lagrangians contributing to electromagnetic masses of $K^0$ will
vanish, then,
\begin{equation}
(m_{K^0}^2)_{EM}=0,\;\;\;\;{in\;\; the\;\; chiral\;\; SU(3)\;\; limit.} 
\end{equation}
Likewise, it can also be sure that the lagrangian (11) and
(12) are exactly
equivalent to lagrangians (9) and (10) respectively under the limit of $
m_\rho=m_\omega=m_\phi$ and $m_a=m_{K_1}$ . Then, according to eqs.(7)
and
(8) we have
\begin{equation}
(m_{K^\pm}^2)_{EM}=(m_{\pi^\pm}^2)_{EM}, \;\;\;\; {in\;\;the\;\;chiral\;\;SU(3)
\;\;limit.} 
\end{equation}

Above arguments and conclusions can also be checked by manifest 
calculations which are standard in quantum fields theory.
From eqs.(9)-(14) and VMD, the electromagnetic masses of $\pi^\pm$,
$K^\pm$ and
$K^0$ can be derived.
To $\pi^\pm$, using the substitution eq.(5) and eqs.(9)(10), we get
${\cal L}_{\gamma\gamma\pi\pi}$,${\cal L}_{\gamma\rho\pi\pi}$
and ${\cal L}_{\gamma\pi\pi a}$. Then, the following 
S-matrices can be calculated
\begin{eqnarray*}
S_\pi(1)&=&\langle \pi|T[i\int d^4x_1{\cal L}_{\gamma\gamma\pi\pi}(x_1)
+{i^2\over 2!}2\int d^4x_1 d^4x_2{\cal L}_{\gamma\rho\pi\pi}(x_1)
{\cal L}_{\rho\gamma}(x_2)\\
&&+{i^3\over 3!}2\int d^4x_1 d^4x_2 d^4x_3
{\cal L}_{\rho\rho\pi\pi}(x_1)
{\cal L}_{\rho\gamma}(x_2){\cal L}_{\rho\gamma}(x_3)] |\pi \rangle, \\
S_\pi(2)&=&\langle \pi|T[
{i^2\over 2!}\int d^4x_1 d^4x_2{\cal L}_{\pi a \gamma}(x_1)
{\cal L}_{\pi a \gamma}(x_2)\\
&&+{i^3\over 3!}6\int d^4x_1 d^4x_2 d^4x_3
{\cal L}_{\pi a \rho}(x_1)
{\cal L}_{\pi a \gamma}(x_2){\cal L}_{\rho\gamma}(x_3) \\
&&+{i^4\over 4!}6\int d^4x_1 d^4x_2 d^4x_3 d^4x_4
{\cal L}_{\pi a \rho}(x_1)
{\cal L}_{\pi a \rho}(x_2){\cal L}_{\rho\gamma}(x_3) 
{\cal L}_{\rho\gamma}(x_4)]|\pi \rangle, 
\end{eqnarray*}
where ${\cal L}_{\rho\gamma}$ has been show in eq.(4). Noting eq.(8)
indicates 
$$
(m_{\pi^\pm}^2)_{EM}={\frac{2i(S_\pi(1)+S_\pi(2))}{\langle \pi |
\pi^2 | \pi \rangle }}, 
$$
then we obtain
\begin{equation}
(m_{\pi^\pm}^2)_{EM} =i{\frac{e^2 }{f_\pi^2}}\int\frac{d^4 k}{(2\pi)^4}
(D-1)m_\rho^4 {\frac{{(F^2+{\frac{k^2 }{2\pi^2}})} }{{k^2 (k^2-m_\rho^2)^2 }}
} [1+{\frac{\gamma^2 }{g^2}}{\frac{{F^2+{\frac{k^2 }{2\pi^2}}} }{{k^2 -m_a^2}
}}]. 
\end{equation}
Similarly, to $K^\pm$ and $K^0$, from eqs.(11)-(14), we have
\begin{eqnarray}
(m_{K^\pm}^2)_{EM}-(m_{K^0}^2)_{EM}&=&i\frac{e^2}{f_\pi^2}\int\frac{d^4 k}{(2\pi)^4}
(D-1)(F^2+\frac{k^2}{2\pi^2})(1+\frac{\gamma^2}{g^2}\frac{F^2
+\frac{k^2}{2\pi^2}}{k^2-m_{K_1}^2})\nonumber \\
& &\times[\frac{1}{3}\frac{m_\rho^2 m_\omega^2}
{k^2(k^2-m_\rho^2)(k^2-m_\omega^2)}+\frac{2}{3}\frac{m_\rho^2 m_\phi^2}
{k^2(k^2-m_\rho^2)(k^2-m_\phi^2)}]
\end{eqnarray}
\begin{eqnarray}
(m_{K^0}^2)_{EM}&=&i\frac{e^2}{4 f_\pi^2}\int\frac{d^4 k}{(2\pi)^4}
(D-1)(F^2+\frac{k^2}{2\pi^2})(1+\frac{\gamma^2}{g^2}\frac{F^2
+\frac{k^2}{2\pi^2}}{k^2-m_{K_1}^2})k^2\nonumber\\
& &\times[\frac{1}{k^2-m_\rho^2}-\frac{1}{3}\frac{1}{k^2-m_\omega^2}-
\frac{2}{3}\frac{1}{k^2-m_\phi^2}]^2
\end{eqnarray}
where $D=4-\varepsilon$.
Obviously, taking $m_\rho=m_\omega=m_\phi$, and $m_a=m_{K_1}$, the contribution
of eq.(21) is zero, and eq.(20) is  exactly eq.(19).  Thus, eq.(18) holds,
and Dashen's theorem(eq.(1)) is automatically obeyed in this theory.

It should be emphasized here that unlike $(m_{\pi^0}^2)_{EM}=0$,
$(m_{K^0}^2)_{EM}=0$ is not only due to the chiral limit, but also
due to the SU(3) symmetry limit for vector mesons.
Similar conclusion has also been obtained by Donoghue and Perez in
Ref.\cite{DP} in the framework of chiral perturbation theory.
Therefore, when one investigates the
violation of Dashen's theorem, the contributions due to
$m_\rho \not= m_\omega \not= m_\phi$ should be taken into account \cite{GYL}.

In the above, the calculations on EM-masses are up to the fourth order
covariant derivatives in effective lagrangians \cite{Li1,Li2}.
In the remainder of this Section,
we find that the result of $(m_{\pi^\pm}^2-m_{\pi^0}^2)_{EM}$ given by
Das et al \cite{DGMLY} can be reproduced if the electromagnetic self-energy of
pions receives the contributions only from the second order derivative terms.
In this case, the interaction lagrangians ${\cal L}_{\rho\rho\pi\pi}$ and
${\cal L}_{\rho\pi a}$ (eqs.(9)(10)) will be simplified as follows
\begin{eqnarray*}
& &{\cal L}_{\rho\rho\pi\pi}=\frac{4F^2}{g^2 f_\pi^2}\rho^0_\mu \rho^{0\mu}
\pi^+\pi^-,\;\;\;\;\;
{\cal L}_{\rho\pi a}=\frac{2iF^2}{f_\pi g^2}\rho^0_{\mu}\pi^+ a^{-\mu}+h.c..
\end{eqnarray*}
Thus, in the chiral limit, the electromagnetic self-energy of pions is
\begin{equation}
(m_{\pi^\pm}^2-m_{\pi^0}^2)_{EM}=(m_{\pi^\pm}^2)_{EM}
=i\frac{3e^2}{f_\pi^2}
\int\frac{d^4k}{(2\pi)^4}m_\rho^4\frac{F^2}{k^2(k^2-m_\rho^2)^2}
(1+\frac{F^2}{g^2(k^2-m_a^2)})
\end{equation}
The Feynman integration in eq.(22) is finite. So it is
straightforward to get the result of $(m_{\pi^\pm}^2-m_{\pi^0}^2)_{EM}$
after performing this integration, which is
\begin{equation}
(m_{\pi^\pm}^2-m_{\pi^0}^2)_{EM}
=\frac{3\alpha_{em}m_\rho^4}{8\pi f_\pi^2}
\{\frac{2F^2}{m_\rho^2}-\frac{2F^4}{g^2(m_a^2-m_\rho^2)}(\frac{1}{m_\rho^2}
+\frac{1}{m_a^2-m_\rho^2}log{\frac{m_\rho^2}{m_a^2}})\}
\end{equation}
where $\alpha_{em}=\frac{e^2}{4\pi}$. Because we only consider the
second order derivative terms in the
lagrangian, the relation between $m_a$ and $m_\rho$ is $m_a^2=\frac{F^2}{g^2}
+m_\rho^2$ instead of eq.(27) in Ref.\cite{Li1}.
Thus, using eq. (15) we can get
\begin{equation}
(m_{\pi^\pm}^2-m_{\pi^0}^2)_{EM}
=\frac{3\alpha_{em}}{4\pi}\frac{m_a^2 m_\rho^2}{m_a^2-m_\rho^2}
log{\frac{m_a^2}{m_\rho^2}}
\end{equation}
When substituting the relation $m_a^2=2 m_\rho^2$, which can be derived from
the Weinberg sum rules \cite{Wb}, into eq.(24), we have
\begin{equation}
(m_{\pi^\pm}^2-m_{\pi^0}^2)_{EM}
=\frac{3log2}{2\pi}\alpha_{em} m_\rho^2
\end{equation}
which is exactly the result obtained by Das et al \cite{DGMLY}, and serves
as the leading term of eq.(19).

\section{Generalization of Dashen's theorem to axial-vector meson
sector}
\par

It is straightforward to extend the studies in the previous section to
the axial-vector octet mesons. In the $U(3)_L\times U(3)_R$ chiral
fields theory of mesons, the lagrangians contributing to electromagnetic
masses of axial-vector mesons ($a_1$ and $K_1$) read
\begin{eqnarray}
{\cal L}_{a a \rho \rho}&=&-\frac{4}{g^2}[
\rho_\mu^0 \rho^{\mu 0} a_\nu^+ a^{-\nu}-\frac{\gamma^2}{2}\rho_\mu^0
\rho_\nu^0(a^{+\mu}a^{-\nu}+a^{-\mu}a^{+\nu})],\\
{\cal L}_{a a \rho}&=&\frac{2i}{g}(1-\frac{\gamma^2}{\pi^2 g^2})
\partial^\nu \rho_\mu^0 a^{+\mu}a_{\nu}^--\frac{2i}{g}\rho_\nu^0 a^{+\mu}
(\partial^\nu a_{\mu}^--\gamma^2 \partial_\mu a^{-\nu})+h.c.,\\
{\cal L}_{a \pi \rho}&=&\frac{2i}{g}(\beta_1 \rho_\mu^0 \pi^+ a^{-\mu}
+\beta_2 \rho_\nu^0 \partial^{\mu\nu}\pi^+ a_{\mu}^-
+\beta_3 \rho_\mu^0 a^{+\mu}\partial^2 \pi^--\beta_4 \rho_\mu^0
\pi^+\partial^2 a^{-\mu}\nonumber \\
& &-\beta_5\rho^{0\mu}\partial_\nu a_{\mu}^+ \partial^\nu \pi^-)+h.c.
\end{eqnarray}
\begin{eqnarray}
{\cal L}_{K_1^+ K_1^- v v}&=&-\frac{1}{g^2}[(\rho_\mu^0+v^8_\mu)^2
K_{1\nu}^+ K_1^{-\nu} \nonumber\\
& &-\frac{\gamma^2}{2}(\rho_\mu^0+v_{\mu}^8)(\rho_\nu^0+v_\nu^8)
(K_1^{+\mu} K_1^{-\nu}+K_1^{-\mu}K_1^{+\nu})], \\
{\cal L}_{K_1^+ K_1^- v}&=&\frac{i}{g}(1-\frac{\gamma^2}{\pi^2 g^2})
(\partial^\nu\rho_{\mu}^0+\partial^\nu v_\mu^8)K_1^{+\mu}K_{1\nu}^{-}
\nonumber \\
& &-\frac{i}{g}(\rho_\nu^0+v_\nu^8)[K_1^{+\mu}(\partial^\nu K_{1\mu}^--\gamma^2
\partial_{\mu}K_{1}^{-\nu})
+ h.c.,  \\
{\cal L}_{K^\pm K_1^\pm v}&=&\frac{i}{g}\beta_1
(\rho_\mu^0+v_\mu^8)K^{+}K_1^{-\mu}
+\frac{i}{g}\beta_2(\rho_\nu^0+v_\nu^8)\partial^{\mu\nu} K^+K_{1\mu}^-\nonumber\\
& &+\frac{i}{g}\beta_3(\rho_\mu^0+v_\mu^8)K_1^{+\mu}\partial^2 K^{-}
-\frac{i}{g}\beta_4(\rho_\mu^0+v_\mu^8)K^{+}\partial^2 K_1^{-\mu}\nonumber \\
& &-\frac{i}{g}\beta_5(\rho_\mu^0+v_\mu^8)\partial_\nu K_{1}^{+\mu}
\partial^\nu K^-
+h.c.
\end{eqnarray}
\begin{eqnarray}
& &{\cal L}_{K_1^0 \bar{K}_1^0 v v}={\cal L}_{K_1^+ K_1^- v v}\{
\rho^0\leftrightarrow -\rho^0,K_1^\pm\leftrightarrow K_1^0(\bar{K}_1^0)\},
\\
& &{\cal L}_{K_1^0 \bar{K}_1^0 v }={\cal L}_{K_1^+ K_1^-  v}\{
\rho^0\leftrightarrow -\rho^0,K_1^\pm\leftrightarrow K_1^0(\bar{K}_1^0)\},
\\
& &{\cal L}_{K_1^0 K^0 v }={\cal L}_{K^\pm K_1^\pm v }\{
\rho^0\leftrightarrow -\rho^0,K_1^\pm\leftrightarrow K_1^0(\bar{K}_1^0)
,K^\pm\leftrightarrow K^0(\bar{K}^0)\}.
\end{eqnarray}
with 
\begin{eqnarray*}
\beta_1=\frac{\gamma F^2}{g f_\pi},\;\;\;
\beta_2=\frac{\gamma}{2\pi^2 g f_\pi},\\
\beta_3=\frac{3\gamma}{2\pi^2 g f_\pi}(1-\frac{2c}{g})+\frac{2\gamma c}{f_\pi},
\;\;\;\beta_4=\frac{\gamma}{2\pi^2 g f_\pi},\\
\beta_5=\frac{3\gamma}{2\pi^2 g f_\pi}(1-\frac{2 c}{g})+\frac{4\gamma c}{f_\pi}.
\end{eqnarray*}
From the above, it is found
that due to the SU(3) symmetry the structure of these axial-vector
meson's lagrangians, eqs.(26)-(34), 
are similar to the pseudoscalar meson's (eqs.(9)-(14)). So it is possible
to find out the EM-masses relations between $a_1$'s and $K_1$'s in the
chiral SU(3) limit 
through similar analyses done in the previous section.
Firstly, like the case of $\pi^0$, there are no couplings to $a_1^0$ field
in lagrangian eqs.(26)-(28), therefore
\begin{equation}
(m_{a^0}^2)_{EM}=0.
\end{equation}
Secondly, according to the previous section, we can take
$v_\mu^8=\rho_\mu^0$ in the calculations of EM-masses of $K_1$ mesons 
in the chiral SU(3) limit.
Then, from lagrangians of eqs.(32)-(34), we have
\begin{equation}
(m_{K_1^0}^2)_{EM}=0.
\end{equation}
Finally, using $v_\mu^8=\rho_\mu^0$ again, noting $m_\pi=m_K=0$ in chiral
limit and comparing $K^\pm$'s lagrangians (eqs.(29)-(31)) with $a_1$'s
(eqs.(26)-(28)), we have
\begin{equation}
(m_{a^\pm}^2)_{EM}=(m_{K^\pm}^2)_{EM}.
\end{equation}
Eqs.(35), (36) and (37) are just eq.(2). Then we conclude that the
generalization of Dashen's theorem to axial-vector mesons is legitimate.

Above conclusion can also be checked by manifest calculations like the case
of pseudoscalar mesons in the previous section. For completeness, we
provide these in follows. The calculations from lagrangians of eqs.(26)-(34)
(together with VMD) to the desired EM-masses are somewhat lengthy, but
the procedures here are clear without
any obstructions, which is like the case of $\pi$'s(eq.(19).

To $a_1$ mesons, the results are as follows,
\begin{equation}
(m_{a^0}^2)_{EM}=0
\end{equation}
\begin{equation}
(m_{a^\pm}^2)_{EM}=(m_{a^\pm}^2)_{EM}(1)+(m_{a^\pm}^2)_{EM}(2)+(m_{a^\pm}^2)_{EM}(3)
\end{equation}
with
\begin{eqnarray}
&&(m_{a^\pm}^2)_{EM}(1)
=ie^2\frac{\gamma^2\langle a| \int d^4 x
a^{\underline{i}\mu}a^{\underline{i}\nu}|a\rangle-\langle a| \int d^4 x
a^{\underline{i}\lambda}a_{\lambda}^{\underline{i}}|a\rangle
g^{\mu\nu}}{\langle a| \int d^4 x a_{\mu}^{\underline{i}}a^{\underline{i}\mu}
|a\rangle}
\nonumber \\
& &\times\int\frac{d^4 k}{(2\pi)^4}\frac{m_\rho^4}{k^2(k^2-m_\rho^2)^2}
(g_{\mu\nu}-\frac{k_\mu k_\nu}{k^2}),\\
\mbox{\vspace{2mm}}
&&(m_{a^\pm}^2)_{EM}(2)
=\frac{ie^2}{\langle a| \int d^4 x a^{\underline{i}\mu}
a_{\mu}^{\underline{i}}|a\rangle}\int\frac{d^4 k}{(2\pi)^4}\frac{1}{k^2-2p\cdot k}
\frac{m_\rho^4}{k^2(k^2-m_\rho^2)^2}\nonumber \\
& &\times\{\langle a| \int d^4 x a^{\underline{i}\mu}a_{\mu}^{\underline{i}}|a\rangle
[4 m_a^2+(b^2+2b\gamma^2)k^2
+2\gamma^4 p\cdot k-\frac{4(p\cdot k)^2}{k^2}\nonumber \\
& &-\frac{1}{m_a^2}(b k^2-(b-\gamma^2)p\cdot k)^2]
+\langle a| \int d^4 x a_{\mu}^{\underline{i}}a_{\nu}^{\underline{i}}|a\rangle
k^\mu k^\nu [-(3b^2-4 b+4)\nonumber \\
& &+D(b+\gamma^2)^2+4\gamma^2
-6b\gamma^2-2\gamma^4-\frac{2\gamma^4 p\cdot k}{k^2}
+\frac{1}{m_a^2 k^2}(b k^2-2(1-\gamma^2)p\cdot k)^2]\},\\
\mbox{\vspace{2mm}}
&&(m_{a^\pm})_{EM}(3)
=\frac{-ie^2}{\langle a| \int d^4 x a_{\mu}^{\underline{i}}
a^{\underline{i}\mu}|a\rangle}\int\frac{d^4 k}{(2\pi)^4}\frac{1}{(p-k)^2}
\frac{m_\rho^4}{k^2(k^2-m_\rho^2)^2}\nonumber\\
& &\times\{\langle a| \int d^4 x a_{\mu}^{\underline{i}}a^{\underline{i}\mu}|a\rangle
(\beta_1^\prime-3 \beta_2 p\cdot k+
\beta_3 k^2)^2+\nonumber \\
& &\langle a| \int d^4 x a_{\mu}^{\underline{i}}a_{\nu}^{\underline{i}}|a\rangle k^\mu k^\nu
[\beta_2 m_{a}^2-\frac{(\beta_1^\prime-2 \beta_2 p\cdot k+\beta_3 k^2)^2}{k^2}
]\}.
\end{eqnarray}
where $\underline{i}$=1,2, $p^2=m_a^2$ and
\begin{eqnarray*}
& &b=1-\frac{\gamma^2}{\pi^2 g^2},\\
& &\beta_1^\prime=\beta_1+(\beta_3+\beta_4-\beta_5) m_{K_1}^2.
\end{eqnarray*}

To $K_1$ mesons, we have
\begin{eqnarray*}
(m_{K_1^\pm}^2)_{EM}-(m_{K_1^0}^2)_{EM}&=&[(m_{K_1^\pm}^2)_{EM}(1)-
(m_{K_1^0}^2)_{EM}(1)]+[(m_{K_1^\pm}^2)_{EM}(2)-(m_{K_1^0}^2)_{EM})(2)]\\
& &+[(m_{K_1^\pm}^2)_{EM}(3)-(m_{K_1^0}^2)_{EM}(3)]
\end{eqnarray*}
with
\begin{eqnarray}
&&(m_{K_1^\pm}^2)_{EM}(1)-(m_{K_1^0}^2)_{EM}(1)
=ie^2\frac{\gamma^2\langle K_1| \int d^4 x
K_1^{\mu+}K_1^{\nu-}|K_1\rangle-\langle K_1| \int d^4 x K_1^{\lambda+} K_{1\lambda}^-|K_1\rangle
g^{\mu\nu}}{\langle K_1| \int d^4 x K_{1\mu}^+K_1^{\mu-}|K_1\rangle}\nonumber \\
& &\times\int\frac{d^4 k}{(2\pi)^4}
(g_{\mu\nu}-\frac{k_\mu k_\nu}{k^2})
[\frac{1}{3}\frac{m_\rho^2 m_\omega^2}{k^2(k^2-m_\rho^2)(k^2-m_\omega^2)}+
\frac{2}{3}\frac{m_\rho^2 m_\phi^2}{k^2(k^2-m_\rho^2)(k^2-m_\phi^2)}],\\
\vspace{2mm}
&&(m_{K_1^\pm}^2)_{EM}(2)-(m_{K_1^0}^2)_{EM}(2)
=\frac{ie^2}{\langle K_1| \int d^4 x K_1^{\mu+}
K_{1\mu}^-|K_1\rangle}\int\frac{d^4 k}{(2\pi)^4}\frac{1}{k^2-2p\cdot k}\nonumber \\
& &\times\{\langle K_1| \int d^4 x K_1^{\mu+}K_{1\mu}^-|K_1\rangle[4 m_{K_1}^2+(b^2+2b\gamma^2)k^2
+2\gamma^4 p\cdot k-\frac{4(p\cdot k)^2}{k^2}\nonumber \\
& &-\frac{1}{m_{K_1}^2}(b k^2-(b-\gamma^2)p\cdot k)^2]
+\langle K_1| \int d^4 x K_{1\mu}^+K_{1\nu}^-|K_1\rangle k^\mu k^\nu [
-(3b^2-4 b+4)\nonumber \\
& &+D(b+\gamma^2)^2+4\gamma^2
-6b\gamma^2-2\gamma^4-\frac{2\gamma^4 p\cdot k}{k^2}
+\frac{1}{m_{K_1}^2 k^2}(b k^2-2(1-\gamma^2)p\cdot k)^2]\}\nonumber \\
& &\times[\frac{1}{3}\frac{m_\rho^2 m_\omega^2}{k^2(k^2-m_\rho^2)(k^2-
m_\omega^2)}+\frac{2}{3}\frac{m_\rho^2 m_\phi^2}{k^2(k^2-m_\rho^2)(
k^2-m_\phi^2)}],\\
\vspace{2mm}
&&(m_{K_1^\pm}^2)_{EM}(3)-(m_{K_1^0}^2)_{EM}(3)
=\frac{-ie^2}{\langle K_1| \int d^4 x K_{1\mu}^+
K_1^{\mu-}|K_1\rangle}\int\frac{d^4 k}{(2\pi)^4}\frac{1}{(p-k)^2-m_K^2}\nonumber\\
& &\times\{\langle K_1| \int d^4 x K_{1\mu}^+K_1^{\mu-}|K_1\rangle(\beta_1^\prime-3\beta_2 p\cdot k+
\beta_3 k^2)^2+\nonumber \\
& &\langle K_1| \int d^4 x K_{1\mu}^+K_{1\nu}^-|K_1\rangle k^\mu k^\nu
[\beta_2 m_{K_1}^2-\frac{(\beta_1^\prime-2 \beta_2 p\cdot k+\beta_3 k^2)^2}{k^2}
]\}\nonumber \\
& &\times[\frac{1}{3}\frac{m_\rho^2 m_\omega^2}{k^2(k^2-m_\rho^2)(k^2-
m_\omega^2)}+\frac{2}{3}\frac{m_\rho^2 m_\phi^2}{k^2(k^2-m_\rho^2)(
k^2-m_\phi^2)}],
\end{eqnarray}
and
$$
(m_{K_1^0}^2)_{EM}=(m_{K_1^0}^2)_{EM}(1)+(m_{K_1^0}^2)_{EM}(2)+(m_{K_1^0}^2)_{EM}(3)
$$
with
\begin{eqnarray}
&&(m_{K_1^0}^2)_{EM}(1)
=ie^2\frac{\gamma^2\langle K_1| \int d^4 x
K_1^{\mu 0}\bar{K}_1^{\nu 0}|K_1\rangle-\langle K_1| \int d^4 x K_1^{\lambda 0}
\bar{K}_{1\lambda}^0|K_1\rangle
g^{\mu\nu}}{4\langle K_1| \int d^4 x K_{1\mu}^0\bar{K}_1^{\mu 0}|K_1\rangle}\nonumber \\
& &\times\int\frac{d^4 k}{(2\pi)^4}
(k^2 g_{\mu\nu}-{k_\mu k_\nu})
[\frac{1}{k^2-m_\rho^2}-\frac{1}{3}\frac{1}{k^2-m_\omega^2}-
\frac{2}{3}\frac{1}{k^2-m_\phi^2}]^2,\\
&&(m_{K_1^0}^2)_{EM}(2)
=\frac{ie^2}{4\langle K_1| \int d^4 x K_1^{\mu 0}
\bar{K}_{1\mu}^0|K_1\rangle}\int\frac{d^4 k}{(2\pi)^4}\frac{1}{k^2-2p\cdot k}\nonumber \\
& &\times\{\langle K_1| \int d^4 x K_1^{\mu 0}\bar{K}_{1\mu}^0|K_1\rangle[4 m_{K_1}^2+(b^2+2b\gamma^2)k^2
+2\gamma^4 p\cdot k-\frac{4(p\cdot k)^2}{k^2}\nonumber \\
& &-\frac{1}{m_{K_1}^2}(b k^2-(b-\gamma^2)p\cdot k)^2]
+\langle K_1| \int d^4 x K_{1\mu}^0\bar{K}_{1\nu}^0|K_1\rangle k^\mu k^\nu [
-(3b^2-4 b+4)\nonumber \\
& &+D(b+\gamma^2)^2+4\gamma^2
-6b\gamma^2-2\gamma^4-\frac{2\gamma^4 p\cdot k}{k^2}
+\frac{1}{m_{K_1}^2 k^2}(b k^2-2(1-\gamma^2)p\cdot k)^2]\}\nonumber \\
& &\times k^2[\frac{1}{k^2-m_\rho^2}-\frac{1}{3}\frac{1}{k^2-m_\omega^2}
-\frac{2}{3}\frac{1}{k^2-m_\phi^2}]^2,\\
\vspace{2mm}
&&(m_{K_1^0}^2)_{EM}(3)
=\frac{-ie^2}{4\langle K_1| \int d^4 x K_{1\mu}^0
\bar{K}_1^{\mu 0}|K_1\rangle}\int\frac{d^4 k}{(2\pi)^4}\frac{1}{(p-k)^2-m_K^2}\nonumber\\
& &\times\{\langle K_1| \int d^4 x K_{1\mu}^0\bar{K}_1^{\mu 0}|K_1\rangle(\beta_1^\prime-3\beta_2 p\cdot k+
\beta_3 k^2)^2+\nonumber \\
& &\langle K_1| \int d^4 x K_{1\mu}^0\bar{K}_{1\nu}^0|K_1\rangle k^\mu k^\nu
[\beta_2 m_{K_1}^2-\frac{(\beta_1^\prime-2 \beta_2 p\cdot k+\beta_3 k^2)^2}{k^2}
]\}\nonumber \\
& & \times k^2[\frac{1}{k^2-m_\rho^2}-\frac{1}{3}\frac{1}{k^2-m_\omega^2}-
\frac{2}{3}\frac{1}{k^2-m_\phi^2}]^2.
\end{eqnarray}
where $p^2=m_{K_1}^2$.

Taking $m_\rho=m_\omega=m_\phi$, and $m_a=m_{K_1}$ (chiral SU(3) limit), 
we immediately obtain
\begin{eqnarray*}
(m_{a^\pm}^2)_{EM}(i)=(m_{K_1^\pm}^2)_{EM}(i),\;\;\;
(m_{K_1^0})_{EM}(i)=0,\;\;\; i=1,2,3.
\end{eqnarray*}
Then the conclusion of eqs.(35)-(37) (eq.(2)) has been reconfirmed.

\section{Generalization of Dashen's theorem to vector meson
sector}
\par

Now let us consider the vector meson octet sector. According to
Refs.\cite{Li1,Li2}, the lagrangians which
contribute to
the electromagnetic masses of $\rho^\pm$ and $K^{*\pm}$ read
\begin{eqnarray}
{\cal L}_{\rho\rho\rho\rho}&=&-\frac{4}{g^2}\rho_\mu^0\rho^{0\mu} \rho_\nu^+
\rho^{-\nu}+\frac{2}{g^2}\rho_\mu^0\rho^0_\nu(\rho^{+\mu}\rho^{-\nu}+
\rho^{-\mu}\rho^{+\nu}),\\
{\cal L}_{\rho\rho\rho}&=&\frac{2i}{g}\partial_\nu \rho_\mu^0\rho^{+\mu}
\rho^{-\nu}-\frac{2i}{g}\rho_\nu^0 \rho_\mu^+(\partial^\nu \rho^{-\mu}-
\partial^\mu \rho^{-\nu})+h.c.,\\
{\cal L}_{\rho\omega\pi}&=&-\frac{3}{\pi^2 g^2 f_\pi}
\epsilon^{\mu\nu\alpha\beta}\partial_\mu\omega_\nu\rho_\alpha^+
\partial_\beta\pi^-+h.c.
\end{eqnarray}
\begin{eqnarray}
{\cal L}_{K^{*+}K^{*-}v v}&=&-\frac{1}{g^2}(\rho_\mu^0+v_\mu^8)^2 K^{+}_\nu
K^{-\nu}\nonumber \\
& &+\frac{1}{2 g^2}(\rho_\mu^0+v_\mu^8)(\rho_\nu^0+v_\nu^8)(K^{+\mu}K^{-\nu}+
K^{-\mu}K^{+\nu}),\\
{\cal L}_{K^{*+}K^{*-} v}&=&\frac{i}{g}(\partial_\nu\rho_\mu^0
+\partial_\nu v_\mu^8) K^{+\mu} K^{-\nu}\nonumber \\
& &-\frac{i}{g}(\rho_\nu^0+v_\nu^8) K^+_{\mu} (\partial_\nu K^{-\mu}-
\partial^\mu K^{-\nu})+h.c.,\\
{\cal L}_{K^{*\pm}K^\pm v}&=&-\frac{3}{\pi^2 g^2 f_\pi}
\epsilon^{\mu\nu\alpha\beta}K^{+}_\mu\partial_\beta K^-
(\frac{1}{2}\partial_\nu\rho_\alpha^0+
\frac{1}{2}\partial_\nu \omega_\alpha+\frac{\sqrt{2}}{2}
\partial_\nu\phi_\alpha)+h.c.
\end{eqnarray}
Eq.(51) and eq.(54) come from the abnormal part of the effective
lagrangian $
{\cal L}_{IM}$(to see Refs.\cite{Li1,Li2}). Thus, similar to the above,
we conclude without any doubt that $
\rho^\pm$ and $K^{*\pm}$  receive the same electromagnetic self-energies
in the chiral SU(3) limit,
\begin{equation}
({m_{\rho^\pm}^2})_{EM}=({m_{K^{*\pm}}^2})_{EM} 
\end{equation}

However, $\rho ^0$ and $K^{*0}$ can also obtain electromagnetic masses even
in the chiral SU(3) limit, which is different from the case of neutral
pseudoscalar and axial-vector mesons. The lagrangian contributing to the
electromagnetic masses of $K^{*0}$ is 
\begin{eqnarray}
& &{\cal L}_{K^{*0} \bar{K}^{*0} v v}={\cal L}_{K^{*+}K^{*-} v v}
\{\rho^0\leftrightarrow -\rho^0, K^{*\pm}\leftrightarrow K^{*0}(
\bar{K}^{*0})\},\\
& &{\cal L}_{K^{*0} \bar{K}^{*0}  v}={\cal L}_{K^{*+}K^{*-}  v}
\{\rho^0\leftrightarrow -\rho^0, K^{*\pm}\leftrightarrow K^{*0}(
\bar{K}^{*0})\},\\
& &{\cal L}_{K^{*0} K^0  v}={\cal L}_{K^{*\pm}K^\pm v}
\{\rho^0\leftrightarrow -\rho^0, K^{*\pm}\leftrightarrow K^{*0}(
\bar{K}^{*0}), K^\pm\leftrightarrow K^0(\bar{K}^0)\}.
\end{eqnarray}
Note that in eq.(58), the combination of the neutral vector mesons is
$-\rho
_\mu +\omega _\mu +\sqrt{2}\phi _\mu $ instead of $-\rho _\mu +\omega _\mu -\sqrt{2
}\phi _\mu $ emerging in eqs.(56)(57) and the lagrangians contributing to
the electromagnetic masses of $K^0$ and $K_1^0$. Therefore, even in the
chiral SU(3) limit, the electromagnetic masses of $K^{*0}$ is nonzero due to
the contribution coming from eq.(58)(the contributions of eqs.(56) and
(57) vanish in
the limit of $m_\rho =m_\omega =m_\phi $).

To electromagnetic masses of $\rho^0$-mesons, the circumstances are more
complicated. The contributions to $(m_{\rho^0}^2)_{EM}$ from ${\cal L}_{IM}$
is 
\begin{equation}
{\cal L}_{\rho\omega\pi}=-\frac{3}{\pi^2 g^2 f_\pi} \epsilon^{\mu\nu\alpha
\beta}\partial_\mu\omega_\nu\rho_\alpha^0 \partial_\beta\pi^0 
\end{equation}
Distinguishing from the case of $K^{*0}$, the direct $\rho^0$-photon
coupling which comes from VMD(eq.(4)) can bring both the tree and one-loop
diagrams contributing to the electromagnetic masses of $\rho^0$ in the
chiral limit. Thus, $(m_{\rho^0}^2)_{EM}$ is also nonzero. Furthermore, from the
point of view of large-$N_c$ expansion, the tree diagrams are $O(N_C)$
while the one-loop diagrams are $O(1)$ \cite{Li1,Gt}. In general, we can not
expect that $(m_{\rho^0}^2)_{EM}= (m_{K^{*0}}^2)_{EM}$ in the chiral SU(3)
limit(only loop diagrams can contribute to $(m_{K^{*0}}^2)_{EM}$).
So the generalization of Dashen's theorem fails for vector meson octet.
Similar result has also been found by Bijnens and
Gosdzinsky \cite{JG}.

\section{Summary and Discussions}
\par

Employing $U(3)_L\times U(3)_R$ chiral fields theory of mesons
and in the chiral SU(3) limit, we have obtained
\begin{eqnarray*}
(m_{\pi^\pm}^2)_{EM}=(m_{K^\pm}^2)_{EM},\;\;\;
(m_{a^\pm}^2)_{EM}=(m_{K_1^\pm}^2)_{EM},\;\;\;
(m_{\rho^\pm}^2)_{EM}=(m_{K^{*\pm}}^2)_{EM},
\end{eqnarray*}
and the electromagnetic masses of $\pi
^0$, $K^0$, $a_1^0$ and $K_1^0$ vanish. Therefore,
Dashen's theorem(eqs.(1) and (2)) holds for both pseudoscalar and
axial-vector mesons in this effective fields theory. However,
the effective lagrangian makes non-zero contributions to
electromagnetic masses of $
\rho ^0$ and $K^{*0}$ even in the chiral SU(3) limit, and VMD yields the
direct coupling of $\rho^0$ and photon(eq.(4)), which provides
other
contributions to the electromagnetic masses of $\rho^0$. Generally, $(m_{\rho
^0}^2)_{EM}\not =(m_{K^{*0}}^2)_{EM}$. Therefore, the generalization of Dashen's
theorem fails for vector-mesons.

Dashen's theorem is valid only in the chiral SU(3) limit. The violation of
this theorem beyond the lowest order has been investigated
extensively \cite{DHW,JB,KK,RU,GYL,Mou}, and a large violation has been revealed in
Refs.\cite{DHW,JB,GYL,Mou}. In particular, a rather large violation of eq.(2)
has been obtained in Ref.\cite{GYL}.

\begin{center}
{\bf ACKNOWLEDGMENTS}
\end{center}
One of the authors(M.L.Y) wishes to thank C.H. Chang for helpful discussions.
This work is partially supported by NSF of China through C.N. Yang.

\end{document}